\documentclass[prapplied,reprint]{revtex4-1}

\draft % marks overfull lines with a black rule on the right
\usepackage{graphicx}
\usepackage{color}
\usepackage{amsmath}

\begin{document}

\newcommand{\vet}[1]{\mathbf{#1}}
\newcommand{\vect}[1]{\mathbf{#1}}
\newcommand{\an}[1]{{\color{black} {#1}}}
\newcommand{\lc}[1]{{\color{red} \bf{#1}}}
\newcommand{\um} {~\mu\mathrm{m}}

% Use the \preprint command to place your local institutional report number
% on the title page in preprint mode.
% Multiple \preprint commands are allowed.
%\preprint{}

\title{A novel setup coupling space-resolved dynamic light scattering and rheometetry unveils the heterogeneous flow field and non-affine dynamics in startup shear of a gel} %Title of paper

% repeat the \author .. \affiliation  etc. as needed
% \email, \thanks, \homepage, \altaffiliation all apply to the current author.
% Explanatory text should go in the []'s,
% actual e-mail address or url should go in the {}'s for \email and \homepage.
% Please use the appropriate macro for the type of information

% \affiliation command applies to all authors since the last \affiliation command.
% The \affiliation command should follow the other information.
\author{A. Pommella}
\email[]{angelo.pommella@umontpellier.fr}
\author{A.-M. Philippe}
\author{T. Phou}
\author{L. Ramos}
\author{L. Cipelletti}

%\homepage[]{Your web page}
%\thanks{}
%\altaffiliation{}
\affiliation{L2C, Univ Montpellier, CNRS, Montpellier, France.}

% Collaboration name, if desired (requires use of superscriptaddress option in \documentclass).
% \noaffiliation is required (may also be used with the \author command).
%\collaboration{}
%\noaffiliation

\date{\today}

\begin{abstract}
We present a new light scattering setup coupled to a commercial rheometer operated in the plate-plate geometry. The apparatus allows the microscopic dynamics to be measured, discriminating between the contribution due to the affine deformation and additional mechanisms, such as plasticity. Light backscattered by the sample is collected using an imaging optical layout, thereby allowing the average flow velocity and the microscopic dynamics to be probed with both spatial and temporal resolution. We successfully test the setup by measuring the Brownian diffusion and flow velocity of diluted colloidal suspensions, both at rest and under shear. The potentiality of the apparatus are explored in the startup shear of a biogel. For small shear deformations, $\gamma \le 2\%$, the rheological response of the gel is linear. However, striking deviations from affine flow are seen from the very onset of deformation, due to temporally and spatially heterogeneous rearrangements bearing intriguing similarities with a stick-slip process.
\end{abstract}

% insert suggested PACS numbers in braces on next line

\maketitle %\maketitle must follow title, authors, abstract and \pacs

% Body of paper goes here. Use proper sectioning commands.
% References should be done using the \cite, \ref, and \label commands

%\section{\label{sec:level1}}
%\subsection{}
%\subsubsection{}

\section{\label{sec:level1}Introduction}

Soft matter is characterized by a complex structure on the nanometer to micron scale, which results in a wide range of dynamical properties and mechanical behavior. This richness is of great academic interest and has far-reaching implications in industrial applications, e.g. in the food, cosmetics, home care, pharmaceutical and packaging industry. Products such as ice creams, gelatin, toothpaste, skin-care creams, detergents, inks and plastics heavily rely on the remarkable properties of foams, emulsions, suspensions, gels, surfactant solutions and polymers.

Rheology is an important characterization tool for soft matter~\cite{larson99,rubinstein03,cantat13} and biological materials~\cite{chen10,wang07}, extensively used not only to quantify the mechanical response of a system, but also to gain insight on its structure and dynamics.
In the last years, there has been a growing interest in coupling rheology to techniques that probe the material structure and dynamics at a microscopic level, such as microscopy and a wide range of scattering methods, from static and dynamic light scattering, diffusing wave spectroscopy, to neutron and X-ray scattering, including X photon correlation spectroscopy~\cite{berne76,weitz_diffusing-wave_1993,mills_third-generation_2002,noauthor_neutron_2002}. The importance of combining a macroscopic rheological investigation with microscopic measurements has been demonstrated in a wide range of problems, from the dynamics of foams~\cite{gopal95,hohler97}, the behavior of colloidal crystals under shear~\cite{molino98,Schall44,reinicke10,torija11,tamborini14}, the heterogeneous flow of  wormlike micelles~\cite{helgeson09,angelico10,ober11,gurnon14}, to the non-affine deformation of polymer gels and glasses~\cite{basu11,nagazi17} and the creep and yielding of dense emulsions and colloidal suspensions~\cite{chen10,hebraud97,petekidis02,bauer06,besseling07,schall07,zausch08,koumakis12,denisov13,knowlton14,sent15,melissa17,aime18}. Simultaneous rheological and microscopic measurements are particularly valuable, since the detailed behavior of complex systems may vary from run to run, especially in the non-linear regime, which is usually the most interesting one. A prototypical example is the wide distribution of breaking times of soft solids loaded at constant stress~\cite{bonn_delayed_1998}.

Although microscopy is unsurpassed in its ability to follow single-particle trajectories, scattering methods have several advantages, which explain their lasting popularity. Scattering experiments are not restricted to specifically tailored particles, as required for real-space particle tracking; they can deal more easily with turbid samples; they afford a larger sample size and thus better statistics; they allow the sample structure and dynamics to be probed over a wider range of length and time scales. The first experiments coupling scattering and rheology focussed on the structure, often probed by small-angle scattering apparatuses that can be relatively easily coupled to a rheometer. Visible light, X-ray radiation and neutrons have been used, leading to the well-established fields of Rheo-SALS~\cite{tolstoguzov74,beysens79,hashimoto86,ackerson88,jan92,lauger95,kume95,vermant98,vermant99,varadan01,scirocco04,gentile12}, Rheo-SAXS~\cite{molino98,ramos04,yang04,bauer06,dicola08,kosaka10,silva10,torija11,ramos11,akkal13,wieland16}, and Rheo-SANS~\cite{porcar02,helgeson09,angelico10,helgeson10,reinicke10,eberle12,lopez12,gentile14,jung14,gurnon14}, respectively.

Dynamic scattering methods coupled to rheology have increasingly become important: various scattering setups have been coupled to commercial rheometers~\cite{hashimoto86,lauger95,vermant99,rusu99,ali16,guo10} or custom-made devices~\cite{erpelding08,nagazi17,aime16}. We shall divide dynamic scattering methods in single-scattering and multiple-scattering techniques. The former include dynamic light scattering~\cite{berne76} (DLS), for visible light, and X-photon correlation spectroscopy~\cite{mills_third-generation_2002} (XPCS), for coherent X rays. Diffusing wave spectroscopy~\cite{weitz_diffusing-wave_1993} (DWS), by contrast, operates in the opposite limit of strong multiple scattering, using laser light as a source. In the context of scattering experiments coupled to rheology, the key difference between single and multiple scattering methods is the fact that the former probe the motion of the scatterers projected on a well-defined direction, while DWS probes displacements isotropically. Indeed, single scattering is only sensitive to the component of the displacement along the direction of the scattering vector $\mathbf{q} = \mathbf{k}_{sc} - \mathbf{k}_{in}$, with $\mathbf{k}_{sc}$, $\mathbf{k}_{in}$ the wave vector of the scattered and incoming radiation, respectively. Therefore, single scattering experiments can be designed so as to probe motion in a specific direction with respect to the relevant direction set by the rheology measurement, e.g. parallel or perpendicular to the flow. This is not possible with DWS, because multiply scattered photons undergo a random walk in the sample, thus propagating in all directions with equal probability.

If the sample is deformed during the measurement, this distinction is crucial: DWS will be sensitive both to the `ideal' part of the displacement field (e.g. the affine deformation field in an ideal solid or the laminar flow field in an ideal fluid), as well as to any additional microscopic dynamics, e.g. due to Brownian motion, reversible non-affine response~\cite{didonna_nonaffine_2005,zaccone_elastic_2013a}, or irreversible, plastic rearrangements. In DWS, disentangling the ideal contribution to that associated to additional (and usually more interesting) dynamics is generally quite difficult~\cite{erpelding08,nagazi17}, except in the simplest cases where both the structure of the flow field and the origin of the additional dynamics are known \textit{a priori}, e.g. for diluted Brownian suspensions under laminar shear flow ~\cite{wu_diffusing-wave_1990}. For this reasons, DWS has been mostly restricted to rheology tests where the sample is macroscopically undeformed, e.g. in stress relaxation~\cite{nagazi_cartographie_2017,kooij_laser_2018}, or to `echo' experiments that probe stroboscopically the sample under a sinusoidal drive, monitoring the microscopic evolution between states that correspond to the same macroscopic deformation~\cite{hebraud97,petekidis02}.

In single scattering experiments, by contrast, these difficulties can be avoided by orienting $\mathbf{q}$ perpendicular to the direction of the deformation field, such that the experiment is only sensitive to deviations of the microscopic dynamics with respect to the behavior of ideal solids or fluids. This approach has been implemented in both XPCS and DLS. In XPCS, it has been used to measure the Brownian dynamics of colloids pumped in a capillary~\cite{busch_dynamics_2008}, a scheme aiming at minimizing the sample exposure to the intense X-ray radiation, which often causes radiation damage. XPCS has also been used to investigate the yielding transition of colloidal gels submitted to an oscillatory shear deformation~\cite{rogers_echoes_2014,leheny_rheo-xpcs_2015}, where the single scattering geometry allowed the authors to study the dynamics in the directions parallel and perpendicular to the flow direction and within each cycle, not just stroboscopically. The ability of resolving non-affine microscopic dynamics thanks to single scattering is also at the core of the small-angle DLS setup described in Refs.~\cite{aime16,aimeDLSshear}, which allowed the microscopic precursors of the macroscopic failure of a colloidal gel to be unveiled in creep experiments~\cite{aime18}.

Both DWS and single scattering methods are usually implemented in the far field, homodyne geometry, where the detector is placed very far from the sample (or, equivalently, in the focal plane of a lens collecting the scattered light), such that only scattered light is detected. Under these conditions, dynamic scattering methods are sensitive to the relative motion of the scatterers, not to their average displacement. This degeneracy is removed in heterodyne experiments, where a static reference beam illuminates the detector, together with the scattered light. The beating between the reference beam and the scattered light causes distinctive oscillations in the measured intensity correlation function, whose time scale is directly related to the scatterers' average velocity. Heterodyne scattering setups coupled to rheology have been implemented both in XPCS~\cite{livet_x-ray_2006,ehrburger-dolle_xpcs_2012}, for studying the relaxation of nanocomposites, and in DLS~\cite{salmon_optical_2003}, to measure the velocity profile in Couette flow.

An alternatively way to measure the (local) flow velocity in rheology experiments has been recently demonstrated~\cite{ali16}, using space-resolved DLS~\cite{duri09,cipelletti13}, also known as photon correlation imaging (PCI). In the setup of Ref.~\cite{ali16}, a 2D detector makes an image of the sample confined in the gap of a Couette cell driven by a commercial rheometer and illuminated by a laser sheet. The image is formed by light scattered at a well defined angle $\theta = 90^\circ$, such that the setup combines features of a traditional scattering experiment and of imaging. The dynamics are observed in the flow-velocity gradient plane. For a sample undergoing a macroscopic deformation, the flow velocity can be measured. Additionally, the relative motion between scatterers can be quantified, but the microscopic dynamics are typically dominated by the contribution of the affine displacement, because this geometry does not allow the scattering vector to be oriented perpendicular to the flow direction. Thus, this apparatus is best suited for characterizing the local flow profile, for samples undergoing a macroscopic deformation, or, alternatively, for measuring the microscopic dynamics for samples macroscopically at rest, e.g. in echo or stress relaxation experiments.

In this paper, we introduce a novel, custom-made dynamic light scattering setup coupled to a commercial rheometer, allowing for the investigation of the microscopic dynamics of driven samples. The setup works in a back-scattering configuration and uses the PCI method to image the flow-vorticity plane in a plane-plane rheological configuration. This allows one to obtain time- and space-resolved information of the sample internal dynamics, decoupling the affine and non-affine contributions to the microscopic dynamics. Additionally, the local flow velocity, averaged over the rheometer gap, can be measured.

The rest of the paper is organized as follows: in Sec.~\ref{sec:level2} we present the experimental setup. In Sec.~\ref{sec:level3} we test it on a simple Newtonian fluid, a diluted Brownian suspension, showing that its microscopic dynamics are correctly captured both at rest and under shear. The local shear rate obtained by optical measurements is successfully compared to that imposed by the rheometer. In Sec.~\ref{sec:level4} we investigate the microscopic dynamics and the mesoscopic deformation of a biogel during shear startup, as an example of the potentiality of the setup. Section ~\ref{sec:level5} concludes the paper with a few final remarks.

\section{\label{sec:level2}Experimental setup and image processing}

\begin{figure*} [htbp]
\includegraphics{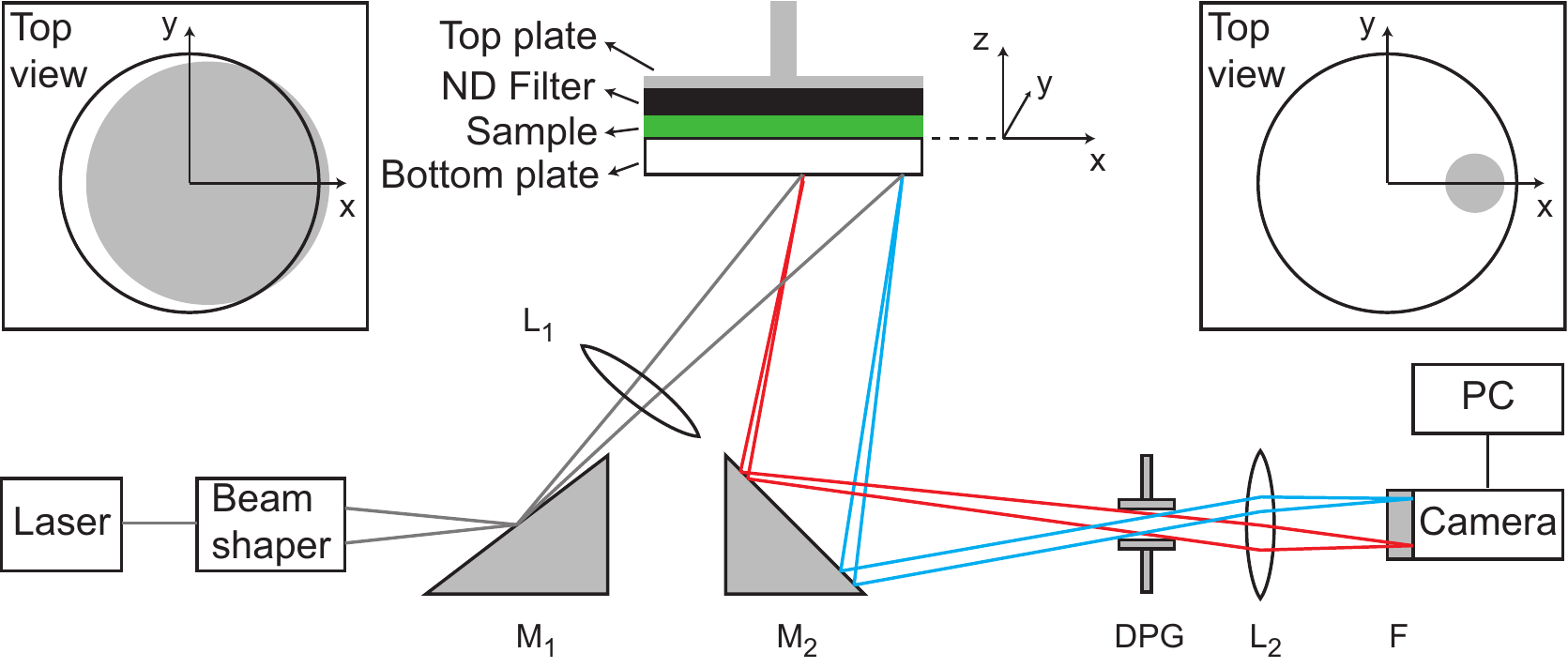}
\caption{\label{Fig1} Scheme of the wide-angle light scattering setup coupled to a rheometer. The incoming laser beam (gray lines) is expanded by a beam shaper. The mirror M$_1$ sends the beam to the sample, which is confined between the bottom plate of the rheometer and a ND filter glued to the top plate%(see Sec.~\ref{sec:level3} for more details)
. The lens L$_1$ controls the size of the illuminated sample. The mirror M$_2$ collects the light back-scattered from the sample (red and blue lines), sending it to the camera. The lens L$_2$ makes an image of the sample on the camera sensor. The iris diaphragm DPG, placed in the focal plane of L$_2$, controls the speckle size. The laser-line filter F cuts ambient light. Images are acquired and processed by the personal computer PC. The optical elements are not to scale. The two insets show as grey disks the size and location with respect to the rheometer plates (white circle) of the largest (left) and smallest (right) illuminated sample region, depending on the choice and positioning of L$_1$. Throughout this paper, we use a reference system with the $(x,y)$ plane at the interface between the bottom plate and the sample and the tool rotation axis as the $z$ axis, oriented upward.}
\end{figure*}

\subsection{Setup}
\label{sec:setup}

Figure~\ref{Fig1} illustrates the experimental setup. A commercial stress-controlled rheometer (Anton Paar MCR 502) is coupled to a custom-made wide-angle light scattering (WALS) apparatus. The rheometer is equipped with a plate-plate geometry with temperature control. The bottom plate is fixed and made of \an{float} glass, to let pass the illuminating laser beam. The top plate, with radius 25 mm, is the rotating one. As discussed in the following, it is vital to minimize any backreflection from the upper plate. To this end, we glue an absorptive neutral density (ND) filter (Absorptive ND Filter by Edmund Optics, optical density 3.0, diameter 50 mm, \an{thickness 3 mm}) to the steel plate (Anton Paar part PP-50) of the rheometer. To allow for an easy replacement of the filter (e.g. when the optical quality of its surface is degraded by use), we use fused saccharose as a glue. The steel plate and the ND filter are placed in an oven at $T = 100^\circ~\mathrm{C}$. The sugar is mixed with a small amount of water and heated so as to obtain a light brown caramel with low viscosity, at $T \approx 160^\circ~\mathrm{C}$. The caramel is poured on the hot plate and the ND filter is glued to the plate. Excess caramel is removed by gently rotating the ND filter. After cooling, a thin layer of nail polish is applied on the rim of the ND-plate assembly, to prevent water to penetrate in the solidified caramel when cleaning the ND filter. To unglue the filter, the nail polish is removed using acetone and the tool is immersed in hot water.

In the plane-plane geometry used here, the shear deformation $\gamma$ and the shear rate $\dot{\gamma}$ vary linearly from zero (at the plate center) to a maximum value (at its edge). For a point at distance $r$ from the rotation axis, one has $\gamma = \Theta r/H$ and $\dot{\gamma} =\omega r / H$, respectively, with $\textit{H}$ the gap between the plates, $\Theta$ the angular rotation of the upper plate, and $\omega = d\Theta/dt$ its angular velocity. Similarly, the shear stress $\sigma$ is not uniform, but rather varies radially as $\sigma = 2 T / \pi r^3$, where $T$ is the torque. Note that, for highly scattering samples (DWS regime), backscattering from the upper tool is not an issue: in this case, one can safely use a conventional plane (glass or metal), or even a cone and plane geometry, which insures constant $\gamma$, $\dot{\gamma}$ and $\sigma$ over the whole sample.

The WALS setup can be divided in two parts. The first part is composed of a laser source and optical elements that shape the illuminating beam. The laser is a single frequency CW diode-pumped surce (Cobolt Samba$^{TM}$ 150, part number 0532-04-01-0150-500), operating at an in-vacuo wavelength $\lambda$ = 532 nm and a maximum power of 150 mW. The collimated beam exiting the laser ($1/e^2$ diameter = 0.7 mm) is first expanded and then focused on the first mirror (M$_1$ in Fig.~\ref{Fig1}) by a beam modifier module, consisting of two diverging lenses (\an{the first plano-concave with} focal length -12.5 mm and \an{and the second bi-concave with focal length} -6.3 mm, \an{both with a diameter of 6.35 mm}) and a converging \an{plano-convex} lens (\an{diameter 25.4 mm}, focal length 62.9 mm). The horizontally-propagating beam is reflected towards the sample by the broadband dielectric mirror M$_1$, tilted by 35$^{\circ}$ with respect to the horizontal plane. Before reaching the sample, the beam passes through the convex lens L$_1$ that allows the size of the illuminated sample to be controlled. Two different \an{bi-convex} lenses are used as L$_1$. A lens with focal length $f_1$ = 38.1 mm \an{and diameter 25.4 mm} is used to illuminate almost the entire 50 mm diameter sample (see the left inset of Fig.~\ref{Fig1}), while a lens with $f_1$ = 100 mm \an{and diameter 50.8 mm} allows a smaller region, \an{of diameter 10-30 mm} to be illuminated, see the right inset of Fig.~\ref{Fig1}.

The second part of the WALS setup forms an image of the illuminated sample onto the sensor of a CMOS camera (acA2000-340km by Basler AG), using the back-scattered light. Note that the directions of the incident and back-scattering light are chosen such as to avoid collecting the specular reflection from the rheometer plates. The sensor has a matrix of 2048 $\times$ 1088 pixels, with a pixel size of 5.5 $\mu$m. A PC equipped with a frame grabber (Solios eV-CLF by Matrox) is used to control the camera and acquire images, through a custom-written software. A laser line filter (F) is placed in front of the camera to reduce the contributions due to ambient light. The backscattered light is first collected by a second broadband dielectric mirror (M$_2$), placed below the sample and oriented at 45$^{\circ}$ with respect to the horizontal plane. The convex lens L$_2$ is used to form the image of the sample on the camera. Two different lenses L$_2$ are used, depending on the size of the illuminated region. When the entire sample is illuminated ($f_1$ = 38.1 mm), a \an{plano-convex} lens with focal length $f_2$ = 85 mm \an{and diameter 50.8 mm} is used, yielding a magnification $M= 0.126$. When illuminating a smaller portion of the sample ($f_1$ = 100 mm), we use a \an{bi-convex} lens with $f_2$ = 200 mm \an{and diameter 25.4 mm}, corresponding to $M=1.422$. An iris diaphragm (DPG) is placed in the front focal plane of L$_2$: its aperture controls the speckle size, which is chosen to be on the order of the pixels size~\cite{viasnoff_multispeckle_2002}. %Details of the various optical and opto-mechanical components of the WALS setup and of their position are provided in the supplementary material \an{ho aggiunto le specifiche delle lenti, direi di cancellarla}.
The whole WALS setup is placed under a box made of black paper, to minimize temperature fluctuations and to protect the setup from ambient light.

In the imaging configuration used here, a given speckle of size $s$ on the sensor results from the interference of light issued from a small sample volume, of later size $s/M$ and depth equal to the gap $H$. By calculating intensity correlation functions averaged on small subsets of the image (regions of interest, ROIs), one can then measure the local dynamics and check for any spatial heterogeneity. The WALS setup is characterized by an average scattering angle $\theta$ = 170$\pm1^{\circ}$ corresponding to a scattering vector $q = 4\pi n\lambda^{-1} \sin (\theta/2) \approx$ 33 $\mu$m$^{-1}$ for water-based samples, with refractive index $n = 1.33$. Note that $q$ varies slightly with the location in the sample, because both the illuminating beam and the collected backscattered light form an $x-$ and $y$-dependent angle with the $z$ axis (see Fig.~\ref{Fig1}). We shall discuss later this dependence; however, we anticipate that in most experiments it has no significant impact on the data analysis.

\subsection{Intensity correlation functions}

Simultaneously to the rheology experiments, a time series of speckle images is acquired either at a constant rate, typically \an{1 to 25} Hz, or using the time-varying scheme of Ref.~\cite{philippe16}. The images are saved on a hard disk for subsequent processing using the Time Resolved Correlation~\cite{cipelletti_time-resolved_2003} method, yielding two-time intensity correlation functions. In brief, the images are corrected for the uneven illumination and dark background as explained in Ref.~\cite{duri_time-resolved-correlation_2005} and a local, two-time degree of correlation $c_I$ is calculated according to
\begin{equation} \label{eq:cI}
c_I(t,\tau,\vet{r}) =  B \frac{\langle I_p(t)I_p(t+\tau) \rangle_{\vet{r}}} {\langle I_p(t) \rangle_{\vet{r}}\langle I_p(t+\tau) \rangle_{\vet{r}}}\,.
\end{equation}
Here, $B \gtrsim 1$ is a normalization factor chosen such that $c_I(\tau \rightarrow 0) = 1$, $I_p(t)$ is the time-dependent intensity measured by the $p$-th pixel, $\tau$ a time delay, and $\langle \cdots \rangle_{\vet{r}}$ is an average over the pixels belonging to a ROI centered around the position $\vet{r} = (x,y)$ (see Fig.~\ref{Fig1} for the choice of the reference system). For stationary dynamics, the previous expression may be averaged over time to improve statistics, yielding the intensity correlation function $g_2-1$:
\begin{equation} \label{eq:g2}
g_2(\tau,\vet{r})-1 =  \left \langle \frac{\langle I_p(t)I_p(t+\tau) \rangle_{\vet{r}}} {\langle I_p(t) \rangle_{\vet{r}}\langle I_p(t+\tau) \rangle_{\vet{r}}} \right \rangle_t \,,
\end{equation}
with $\langle \cdots \rangle_{t}$ the average over time. The intensity correlation function is directly related to the intermediate scattering function $f(\vet{q},\tau)$ (ISF) by the Siegert relation~\cite{berne76}, $g_2-1 = f^2$. The ISF quantifies the microscopic dynamics projected onto the scattering vector:
\begin{equation}
f(\vet{q},\tau) = \frac {\left<\sum_{j,k} \exp[-i\vet{q}\cdot(\vet{r}_j(0)-\vet{r}_k(\tau))]\right>}{\left<\sum_{j,k} \exp[-i\vet{q}\cdot(\vet{r}_j(0)-\vet{r}_k(0))]\right>} \,,
\label{eqn:ISF}
\end{equation}
where the double sum runs over all particles in the scattering volume associated to the analyzed ROI and the brackets indicate an ensemble average.  In order to achieve sufficient statistics, the ROIs must contain a large enough number of speckles: depending on $M$, the minimum lateral size of the associated sample volume typically ranges from \an{$185 \um$ to about 5 mm}, setting the level of coarse graining with which local dynamics are measurable.

\section{\label{sec:level3}Newtonian sample}

\subsection{\label{sec:brownian_rest}Brownian suspension at rest}

We first test our setup on a diluted suspension of colloidal particles, for which we expect Brownian dynamics at rest and Newtonian rheological behavior under shear. The suspending solvent is a mixture of 40\% (v/v) MilliQ water and 60\% (v/v) polyoxyalkylene glycol (Emkarox HV 45-LQ-CQ, Croda Chocques SAS), for which we measure a Newtonian viscosity $\eta$ = 5.5 Pa s. Two kinds of particles were used, at a volume fraction of 0.004\%, so as to be in the single scattering regime. Polystyrene particles with diameter $2R = 105\pm 4~\mathrm{nm}$ were obtained from Micro Particles GmbH, while fluorescent sulfate polystyrene particles with $2R = 190\pm 4~\mathrm{nm}$ were obtained from Molecular Probes. Note that the fluorescent emission is cut by the laser-line filter placed in front of the camera, such that our experiments are only sensitive to scattered light, not fluorescent light.

\begin{figure}
\includegraphics{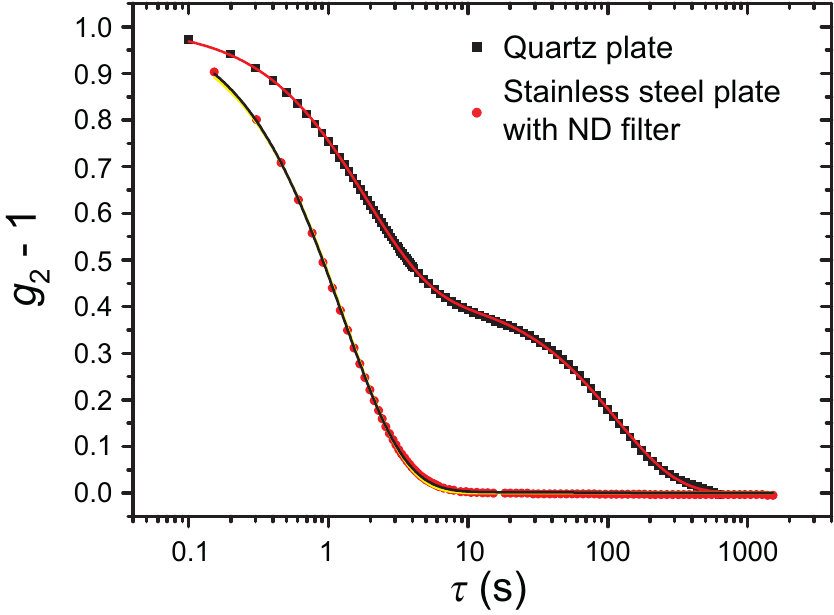}
\caption{\label{Fig2} Intensity correlation functions for a diluted suspension of Brownian particles ($2R$ = 190 nm) at rest. Black squares and red circles are data obtained using as the top rheometer tool a transparent quartz plate or an absorptive ND filter glued to a stainless steel plate, respectively. The lines are fits to the data with a double exponential decay, Eq.~\ref{eqn:g2_2modes}. Two distinct modes are seen with the transparent plate, while the expected single exponential decay is recovered with the absorptive filter, as discussed in the text.}
\end{figure}

\begin{figure}
\includegraphics{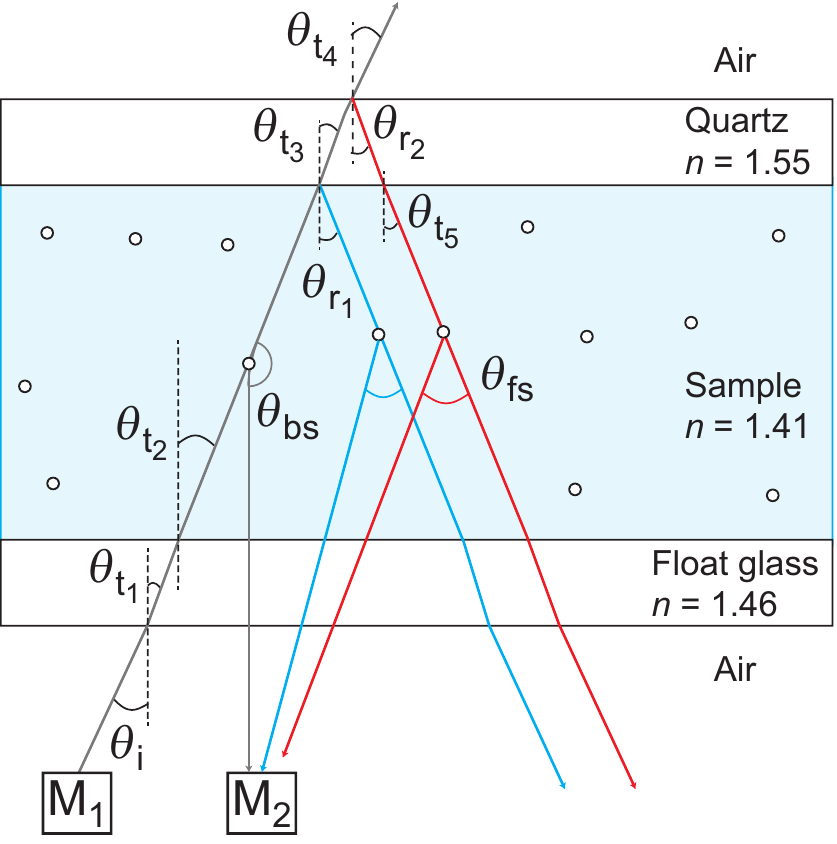}
\caption{\label{Fig3} Scheme of the incoming beam impinging on the bottom plate, passing through the sample and hitting the upper plate. The mirror M$_1$ sends upward the laser beam. The mirror M$_2$ collects light backscattered by particles illuminated by the up-propagating beam (in grey), as well as light scattered forward by particles illuminated by the blue and red beams, which originate from the back reflection of the incoming beam at the sample-upper plate and upper plate-air interfaces. The scheme is not to scale: $\theta_{fs}$ is essentially the same for the beams shown in blue and red. The values of the various angles shown in the figure are: $\theta_i$ = 10$\pm1^{\circ}$, $\theta_{t_1}$ = 6.8$\pm1^{\circ}$, $\theta_{t_2} =\theta_{t_5} = 7.1~\pm 1^{\circ}$, $\theta_{t_3}$ = 6.4$\pm1^{\circ}$, $\theta_{t_4}$ = 10$\pm1^{\circ}$, $\theta_{r_1}$ = 7.1$\pm1^{\circ}$, $\theta_{r_2}$ = 6.4$\pm1^{\circ}$, $\theta_{bs}$ = 170$\pm1^{\circ}$, $\theta_{fs}$ = 10$\pm1^{\circ}$.}
\end{figure}

We first discuss the dynamics at rest. The black squares in Fig.~\ref{Fig2} show $g_2-1$ for the particles with $2R = 190~\mathrm{nm}$, measured using a transparent quartz upper plate (PP43/GL-HT by Anton Paar). Surprisingly, the correlation function exhibits a two-step decay. This is in stark contrast with expectations for a diluted Brownian suspension, for which~\cite{berne76} $f(q,\tau) = \exp(-Dq^2\tau)$, with $D = k_BT/(6\pi \eta R)$ the particle diffusion coefficient, $k_B$ Boltzmann's constant and $T$ the absolute temperature. Tests with different kinds of particles and upper plates suggest that the slower relaxation mode may be due to the (partial) reflection of the incoming beam at the sample-upper plate and upper plate-air interfaces (see Fig.~\ref{Fig3}). These reflected beams propagate back in the sample, illuminating the particles. As a consequence, the CMOS camera receives both light backscattered at an angle $\theta_{bs}$ (from particles illuminated by the upward propagating incident beam, shown in grey in Fig.~\ref{Fig3}) and light forward-scattered at an angle $\theta_{fs}$ (from particles illuminated by the downward propagating back-reflected red and blue beams of Fig.~\ref{Fig3}). Since the reflection coefficient is of the order of a few percents, the intensity of the downward propagating back-reflected beam is much smaller than that of the primary incoming beam. One might then think that the forward scattering contribution should be negligible. However, colloidal particles with diameter of a hundred of nm or more scatter light much more efficiently in the forward direction than in backscattering, thereby compensating for the smaller power of the down-propagating illuminating beam.

To quantitatively test this hypothesis, we model the intensity correlation function as the result of two independent contributions, associated with backscattering and forward scattering from Brownian particles:
\begin{equation}
g_2(\tau)-1 = [A_{bs}\exp(-Dq_{bs}^2\tau) + A_{fs}\exp(-Dq_{fs}^2\tau)]^2 \,,
\label{eqn:g2_2modes}
\end{equation}
where the $bs$ and $fs$ indexes refer to backscattering and forward scattering, respectively, $q_{bs}$ and $q_{fs}$ are the scattering vectors associated to $\theta_{bs}$ and $\theta_{fs}$, and $A_{bs}$ and $A_{fs}$ are the relative weights of the two contributions, with $A_{bs}+A_{fs}=1$. The relative weight $A_{\alpha}$, with $\alpha\in\{bs, fs\}$, is proportional to the scattered intensity of that mode, hence to the intensity of the illuminating beam times the particle form factor $P(q_{\alpha})$. Accordingly, the ratio of the mode weights reads
\begin{equation}
\frac{A_{fs}}{A_{bs}} = \frac{P(q_{fs})\left[R_p^{s-q}+(1-R_p^{s-q})R_p^{q-a}(1-R_p^{q-s})\right ]}{P(q_{bs})} \,,
\label{eqn:reflectivity}
\end{equation}
where we indicate by $R^{\alpha}_p = \mid \frac{n_1 \cos \theta_\an{t} - n_2 \cos \theta_\an{i}}{n_1 \cos \theta_\an{t} + n_2 \cos \theta_\an{i}} \mid^2$ the power reflectivity of light impinging on the interface between two media of refractive indexes $n_1$, $n_2$, with $\theta_i$ and $\theta_t$ the angles of the incident and \an{transmitted} rays to the normal of the interface~\cite{born2013principles}. In Eq.~\ref{eqn:reflectivity}, the subscript $p$ indicates that in our experiment the polarization is that of the $p$ wave~\cite{born2013principles}, while the superscript $\alpha$ indicates the interfaces: $\alpha \equiv s-q$ refers to the sample-quartz plate interface and similarly for the other superscripts, $a$ standing for air. The first term in the brackets of the r.h.s. of Eq.~\ref{eqn:reflectivity} is the contribution due to the light reflected at the $s-q$ interface (blue beam in Fig.~\ref{Fig3}), while the product of the three subsequent factors accounts for light that penetrates in the quartz, is reflected at the $q-a$ interface, and is finally transmitted through the $q-s$ interface back into the sample (red beam in Fig.~\ref{Fig3}).

We fit the data taken with the quartz upper plate with the double exponential decay of Eq.~\ref{eqn:g2_2modes}, using $t_{bs,fs} \equiv 1/Dq_{bs,fs}^2$ and $A_{bs,fs}$ (with $A_{bs}+A_{fs} = 1$) as fitting parameters. As shown by the red solid line in Fig.~\ref{Fig2}, this expression reproduces very well the data. From the measured values of the viscosity and $T$ and using $2R = 190~\mathrm{nm}$ as provided by the manufacturer, we calculate $D =  3.78 \times 10^{-4}\um^2~\mathrm{s}^{-1}$. Using the setup geometrical parameters and the sample refractive index $n_s = 1.41$, we obtain $q_{bs} = 33.18 \pm 0.05 \um^{-1}$, which combined with $D$ yields an expected relaxation time $t_{bs} = 2.40 \pm 0.01~\mathrm{s}$, in very good agreement with $t_{bs} = 2.2~\mathrm{s}$ as obtained from the fit. The relaxation time of the slower mode issued from the fit is $t_{fs} = 233.2~\mathrm{s}$. Using the aforementioned value of $D$, we obtain $q_{fs} = 3.37\um^{-1}$, implying $\theta_{fs} = 11.6^{\circ}$, in good agreement with $\theta_{fs} = 10^{\circ}\pm 1^{\circ}$ as calculated form the setup geometry. The relative amplitude of the two modes is also in fair agreement with the predictions of our simple model: from the fit we obtain $A_{fs}/A_{bs} = 0.53$, to be compared to $0.62 \pm 0.01$ obtained from Eq.~\ref{eqn:reflectivity}, using the values of the refractive indexes and angles given in the caption of Fig.~\ref{Fig3} and $P(q_{fs})/P(q_{bs}) = 13.23$, as obtained from Mie scattering theory~\cite{Kerkerscatteringlightother1969} using the free package MiePlot~\cite{laven_mieplot_}. Note that $R_p^{s-q} =0.22\% \ll R_p^{a-q} = 4.47\%$, such that the most important contribution to forward scattering is that due to the reflection at the upper plate-air interface (red ray in Fig.~\ref{Fig3}), rather than at the sample-plate interface.

Our analysis explains well the double relaxation observed for a model suspension of Brownian particles. For the general case of an arbitrary sample with unknown dynamics, the presence of the slower mode greatly complicates the data analysis, since in general neither its relative amplitude nor its time scale can be easily predicted. We therefore modify the setup in order to make the contribution of forward scattering negligible. As explained in Sec.~\ref{sec:setup}, this is achieved by replacing the quartz upper plate by a neutral density (ND) absorptive filter, with refractive index $n_{ND} = 1.514$, glued to a conventional stainless steel plate (Anton Paar PP-50). Because the refractive index of the ND filter is closer to that of the sample as compared to quartz, this reduces the reflection coefficient at the sample-plate interface to $R_p^{s-ND} = 0.12\%$. Even more importantly, the filter has an optical density of 3 and thus it reduces the intensity of the ray shown in red in Fig.~\ref{Fig3} by a factor~$\gtrsim (10^{-3})^2$, virtually eliminating the dominant contribution to forward scattering.

Figure~\ref{Fig2} shows as red circles $g_2-1$ measured for the same Brownian suspension, with the ND filter glued to the upper plate. Clearly, the slow mode has been almost completely suppressed. Indeed, fitting with a double exponential decay (Eq.~\ref{eqn:g2_2modes}, black line) or by imposing a single mode (Eq.~\ref{eqn:g2_2modes} with $A_{fs}$ set to zero, yellow line) yields almost indistinguishable results. For the two-modes fit, we impose $t_{fs} = 233.2~\mathrm{s}$ as in the experiment without the ND filter, finding $t_{bs} = 2.4~\mathrm{s}$ and $A_{fs}/A_{bs}$ = 0.04. The fitted relaxation time is in excellent agreement with the expected one, while the ratio of the amplitude modes is in fair agreement with $A_{fs}/A_{bs} = 0.016$, as calculated from Eq.~\ref{eqn:reflectivity}, replacing $R_p^{s-q}$ by $R_p^{s-ND} = 0.12\%$ and neglecting the subsequent terms, due to the strong attenuation of the ND filter.
Fitting with a single exponential yields $t_{bs} = 2.6~\mathrm{s}$, very close to the expected value. This demonstrates that  the ND filter is an effective and practical remedy against artifacts due to back-reflections: in most experiments the residual contribution due to forward scattering can be neglected and data can be analyzed by assuming pure backscattering conditions. In the following, we will follow systematically this simpler approach.

\subsection{\label{sec:brownian_shear}Sheared Brownian suspension}

Having validated the measurement of the microscopic dynamics of a sample at rest, we now discuss the case where the sample is sheared at a constant rate $\dot{\gamma}$, using diluted suspensions of Brownian particles as reference systems to test our theoretical analysis. In general, under shear and in the imaging geometry used here, the speckle pattern evolves as a result of three different mechanisms: i) each speckle is advected due to the circular motion imposed by the rheometer; ii) the speckle intensity changes due to the relative motion of the scatterers resulting from the imposed affine deformation field; iii) the speckle intensity fluctuates due to any additional source of microscopic dynamics, e.g. Brownian motion and any non-affine displacement induced by the shear. The first contribution leads to a decay of $g_2-1$ on the time scale it takes a speckle to drift over a distance comparable to its size~\cite{cipelletti13}. For a speckle at a distance $r$ from the rotation axis, this time scale depends on the speckle size, controlled by the magnification $M$ and the aperture of the diaphragm DPG in Fig.~\ref{Fig1}, and on the average drift velocity $\omega r/2$, where the factor of $1/2$ is due to the fact that the lower plate is immobile, while the upper one rotates at an angular speed $\omega$. As we shall discuss it later, the drift contribution can be corrected for by using mixed spatio-temporal intensity correlation functions that probe the dynamics in a reference system co-moving with the sample midplane~\cite{cipelletti13}. For the sake of simplicity, however, we start by considering the case where the affine deformation and the microscopic dynamics induce a decay of $g_2-1$ much faster than that due to advection, such that only the mechanisms ii) and iii) above need to be considered.

%%%%%%%%%%%%%%%%%%%%%%%%%%%%%%%%%%%%%%%%%%%%%%%%%%%%%%%%%%%%%%%%%%%%%%%%%%%%%%%%%
\begin{figure}
\includegraphics{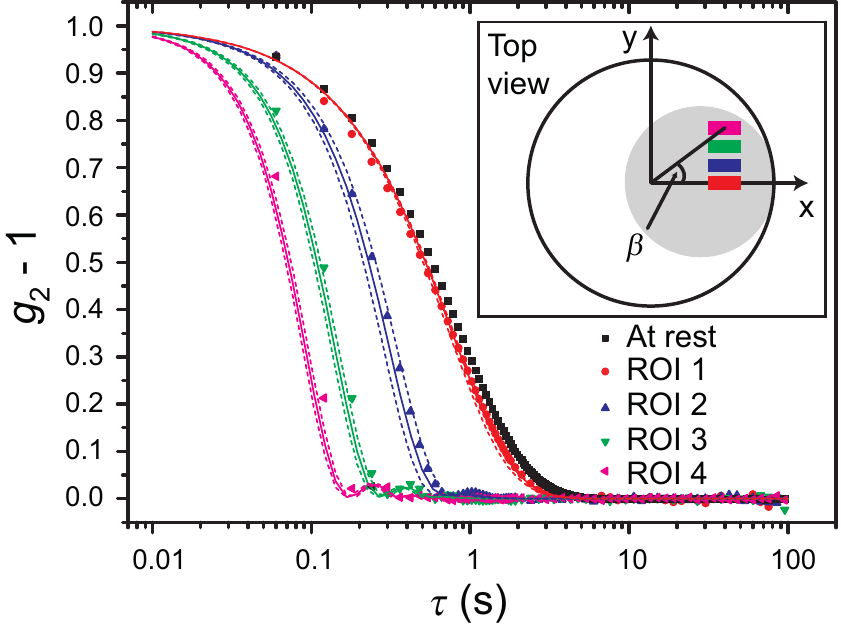}
\caption{\label{Fig4}
Symbols: intensity correlation functions measured in four ROIs, for a sheared suspension of Brownian particles with $D = 5.7 \times 10^{-4}\um^2 \mathrm{s}^{-1}$. %Brownian particles with $D_p$ = 105 nm are used.
The ROI positions in the $x-y$ plane are shown in the inset, where the gray circle represents the portion of the sample illuminated by the laser beam. In the main graph, data for the same sample at rest are shown as black squares. The solid lines are the behavior expected from the combination of Brownian motion and affine displacements, Eq.~\ref{eq:g2_shear_brownian}, with no fit parameters. The dotted lines show the confidence band of the theoretical $g_2-1$, due to the uncertainty on $q_x$ and $q_y$. For all ROIs, $q = 33.2 \um^{-1}$, while the parameters used in Eq.~\ref{eq:g2_shear_brownian} to account for affine displacements are as follows: ROI 1: $q_x$ = -3.13$\pm$0.09 $\mu$m$^{-1}$, $q_y$ = 0 $\mu$m$^{-1}$, $\beta$ = 0$^{\circ}$, $\dot{\gamma}$ = 1.36 $\times$ 10$^{-2}$ s$^{-1}$. ROI 2: $q_x$ = -3.02$\pm$0.09 $\mu$m$^{-1}$, $q_y$ = 0.27$\pm$0.09 $\mu$m$^{-1}$, $\beta$ = 21.5$^{\circ}$, $\dot{\gamma}$ = 0.98 $\times$ 10$^{-2}$ s$^{-1}$. ROI 3: $q_x$ = -3.02$\pm$0.09 $\mu$m$^{-1}$, $q_y$ = 0.612$\pm$0.09 $\mu$m$^{-1}$, $\beta$ = 44$^{\circ}$, $\dot{\gamma}$ = 1.27 $\times$ 10$^{-2}$ s$^{-1}$. ROI 4: $q_x$ = -3.02$\pm$0.09 $\mu$m$^{-1}$, $q_y$ = 0.96$\pm$0.09 $\mu$m$^{-1}$, $\beta$ = 57$^{\circ}$, $\dot{\gamma}$ = 1.68 $\times$ 10$^{-2}$ s$^{-1}$.}
\end{figure}
%%%%%%%%%%%%%%%%%%%%%%%%%%%%%%%%%%%%%%%%%%%%%%%%%%%%%%%%%%%%%%%%%%%%%%%%%%%%%%%%%%%

Figure~\ref{Fig4} shows the intensity correlation functions measured for a diluted suspension of Brownian particles with $2R = 105$ nm, for four different ROIs, while the sample is sheared by rotating the upper plate at a constant angular velocity $\omega$ = 1.26 $\times$ 10$^{-3}$ rad/s, with a gap fixed to $H = 1.05~\mathrm{mm}$. The case at rest is also reported for comparison. The location of the ROIs is sketched in the inset, their size is $50 \times 10$ pixel$^2$, corresponding to $2.2 \times 0.4~\mathrm{mm}^2$ in the sample. Clearly, shearing the sample results in a faster decay of $g_2-1$. The acceleration of the dynamics depends strongly on the location of the ROI: it is negligible for ROI 1, located on the $x$ axis, while the decay rate grows up to more than a factor of 10 for ROI 4, which has the largest $y$ component.

In order to understand this behavior, we model the decay of the intermediate scattering function $f$, Eq.~\ref{eqn:ISF}, for a sample under shear. Neglecting for the moment the contribution due to the average advection [i) above], we assume that the displacements due to the microscopic dynamics and to the affine deformation field are uncorrelated, which results in the factorization
\begin{equation} \label{eq:factorization}
f(\mathbf{q},\tau) = f_{\mu}(\mathbf{q},\tau)f_{aff}(\mathbf{q},\tau)\,,
\end{equation}
where the subscripts $\mu$ and $aff$ refer to the microscopic dynamics and affine deformation contributions, respectively. For a Brownian suspension, the former is the same as for the unperturbed sample discussed in reference to Fig.~\ref{Fig2}, $f_{\mu}(\mathbf{q},\tau) = \exp(-Dq^2\tau)$. The latter can be evaluated following the approach of Ref.~\cite{aimeDLSshear}. For a Newtonian suspension, the shear flow is purely affine, with no $z$ component, such that the affine displacement of a particle with coordinates $(x,y,z)$ over a time $\tau$ is $\Delta_{aff} \mathbf{r}(\tau) = \omega z \tau \sqrt{x^2+y^2}/H$.

We start by considering particles laying on the $x$ axis; later, we shall generalize our results to particles with arbitrary $(x,y)$ coordinates. For particles with $y=0$, $\Delta_{aff}\mathbf{r}(\tau) = \dot{\gamma}(x,y) z \tau \hat{u}_y$, where we have introduced the local shear rate $\dot{\gamma}(x,y) = \omega \sqrt{x^2+y^2}/H$ and where $\hat{u}_y$ is the unit vector along the $y$ axis. The affine contribution to the intermediate scattering function then reads
\begin{equation}
\begin{split}
\label{eq:affine_x_sum}
& f_{aff} (\vect{q},\tau) = \\
& \frac{1}{N} \left \langle \sum_{j=1}^{N} e^{-i\mathbf{q} \cdot \Delta_{aff}\mathbf{r}_j (\tau)} + \sum_{j \neq l=1}^{N} e^{-i\mathbf{q} \cdot [\mathbf{r}_j (t)-\mathbf{r}_l (t+\tau)]} \right \rangle \,.
\end{split}
\end{equation}
If the particle positions are uncorrelated, as in our Brownian suspension, the second sum vanishes. By replacing the average and the first sum by an integral of the exponential term weighted by the (flat) probability distribution function of the particle position~\cite{berne76}, one finds
\begin{equation}
\begin{split}
\label{eq:affine_x}
&f_{aff} (\vect{q},\tau) = \\
&\int_{0}^{H} e^{-i q_y \dot{\gamma} z \tau} dz = \mathrm{sinc} \left( \mathbf{q} \cdot \hat{u}_y \frac {\dot{\gamma} H \tau} {2}\right) e^{i \mathbf{q} \cdot \hat{u}_y  \dot{\gamma} \frac{H}{2} \tau}\,.
\end{split}
\end{equation}

The case of particles with arbitrary $(x,y)$ coordinates is simply obtained from Eq.~\ref{eq:affine_x}. By replacing $\hat{u}_y$ by the general expression for the unit vector parallel to the direction of the flow, $\sin \beta \hat{u}_x + \cos\beta \hat{u}_y$, where the angle $\beta$ is defined in Fig.~\ref{Fig4}, one finds
\begin{equation}
\begin{split}
\label{eq:affine}
f_{aff} (\vect{q},\tau)  & = \mathrm{sinc} \left [ \left( q_x \sin\beta + q_y\cos \beta  \right)\frac {\dot{\gamma} H \tau} {2}\right] \\
& \times \exp\left[i \left( q_x \sin\beta + q_y\cos \beta  \right) \dot{\gamma} \frac{H}{2} \tau\right]\,.
\end{split}
\end{equation}
Equation~\ref{eq:affine} shows that $f_{aff}$ depends only on the $x$ and $y$ components of $\mathbf{q}$. This is a consequence of the fact that DLS probes particle displacements projected onto the direction of the scattering vector and that affine displacements occur in the $(x,y)$ plane. Although in our setup the $z$ component of $\mathbf{q}$ is the largest one (typically, $|q_z| \approx 33 \um^{-1}$ and $|q_x|,~|q_y| \lesssim 3 \um^{-1}$), the contribution of motion in the $(x,y)$ plane cannot be neglected, due to the coupling with the shear flow.

Using the Siegert relationship, Eq.~\ref{eq:affine}, and the factorization of Eq.~\ref{eq:factorization}, we finally obtain the following expression for the intensity correlation function of a sheared Brownian suspension:
\begin{equation}
\begin{split}
\label{eq:g2_shear_brownian}
& g_2 (\mathbf{q},\dot{\gamma}(x,y),\tau ) - 1 = \exp(-2D q^2 \tau)\\
& \times \mathrm{sinc}^2\left[\frac{\dot{\gamma}(x,y) \tau H} {2}(q_x \sin \beta + q_y \cos \beta)\right] \,.
\end{split}
\end{equation}
Note that in writing Eq.~\ref{eq:g2_shear_brownian} we have assumed that the shear does not induce any additional dynamics, besides that due to the affine flow field. This is justified for diluted suspensions, but may not hold at higher concentrations, where hydrodynamic interactions cannot be neglected.

We now use Eq.~\ref{eq:g2_shear_brownian} to model the experimental correlation functions shown as symbols in Fig.~\ref{Fig4}. To account for the finite size of the ROIs, Eq.~\ref{eq:g2_shear_brownian} is integrated over the $(x,y)$ extension of each ROI, using the nominal values of all parameters. This yields the solid lines in Fig.~\ref{Fig4}; the dotted lines show the confidence band for the theoretical $g_2-1$ due to the uncertainty on $q_x$ and $q_y$. An excellent agreement is found between the data and  Eq.~\ref{eq:g2_shear_brownian}, with no fitting parameters, thus demonstrating the validity of our theoretical analysis. For ROI 1, $g_2-1$ is very close to the correlation function for the same sample at rest, because for that ROI $q_y \approx 0$ and $\sin \beta \approx 0$, such that the affine displacement projected onto the scattering vector almost vanishes. The slightly faster decay of the data under shear with respect to the sample at rest is due to the finite size of ROI 1 and the finite range of scattering angles accepted by the collection optics, which implies that $g_2-1$ contains contributions of pixels associated to small yet finite $q_y$ and $\sin \beta$ values. While for ROI 1 Brownian motion is overwhelmingly responsible for the decay of $g_2-1$, for the other ROIs the contribution of affine motion is the dominant one. The data shown in Fig.~\ref{Fig4} demonstrate that the contribution of the (usually uninteresting) affine deformation may or may not be relevant, depending on the ROI location and the relative importance of the microscopic dynamics with respect to affine displacements. Quite importantly, Eq.~\ref{eq:g2_shear_brownian} allows the decay rate due to affine motion to be reliably predicted, thus providing a means to identify any additional dynamics, e.g. due to plastic rearrangements. As a final comment, it is worth mentioning that for practical purposes the integration over the ROI may be avoided by using directly  Eq.~\ref{eq:g2_shear_brownian} and letting $\beta$ as a free parameter. The $\beta$ value thus obtained represents an `effective' angular position of the ROI that nicely accounts for its finite size. For example, we find that the continuous line calculated by integration for ROI 1 is virtually indistinguishable (maximum difference $< 5 \times 10^{-3}$) from Eq.~\ref{eq:g2_shear_brownian} evaluated for the center of the ROI and an effective $\beta = 1.6^\circ$. Note that the effective $\beta$ is intermediate between $\beta = 0$ (at the ROI center) and the maximum value $\beta = 2.7^\circ$ (at the ROI top left corner).

In Fig.~\ref{Fig4} the applied rotation speed was quite low, resulting in a modest drift velocity of the speckles, ranging from 0.3 pixels/s for ROI 1 to 0.4 pixels/s for ROI 4. Because the affine deformation and Brownian motion induce a decay of $g_2-1$ on the time scale of about 1 sec (for ROI 1) or much smaller (for all other ROIs), the contribution of speckle advection is negligible. In other words, the speckle intensity fluctuates much faster than the time it takes a speckle to be advected over a distance of one pixel, thus making negligible the contribution of mechanism i) discussed at the beginning of Sec.~\ref{sec:brownian_shear}. We now explore the opposite limit of a Brownian suspension sheared at a much higher rate, for which advection cannot be neglected.

The method for correcting for speckle drift has been described in detail in Ref.~\cite{cipelletti13}. In brief, a spatio-temporal degree of correlation that generalizes Eq.~\ref{eq:cI} is calculated:
\begin{equation} \label{eq:cI_spatio-temporal}
c_I(t,\tau,\vet{r},\Delta x, \Delta y) =  B \frac{\langle I_p(t)I_{p,\Delta x, \Delta y}(t+\tau) \rangle_{\vet{r}}} {\langle I_p(t) \rangle_{\vet{r}}\langle I_{p,\Delta x, \Delta y}(t+\tau) \rangle_{\vet{r}}}\,,
\end{equation}
where the subscripts $p,\Delta x, \Delta y$ indicate a pixel spatially shifted by $(\Delta x, \Delta y)$ with respect to the location of pixel $p$. At fixed $\tau$, the spatio-temporal degree of correlation typically exhibits a peak as a function of the spatial lag, whose position $(\Delta x^*, \Delta y^*)$ provides the speckle drift between time $t$ and $t+\tau$. The height of the peak, $c_I^* = c_I(t,\tau,\vet{r},\Delta x^*, \Delta y^*)$, represents the degree of correlation corrected for the effect of drift, e.g. the loss of correlation due only to the relative motion of the scatterers. For a stationary drift motion, as in the experiments reported here, $c_I^*$ is averaged over time, yielding an intensity correlation function $g_2-1$ corrected for the drift contribution. Note that the method introduced in Ref.~\cite{cipelletti13} assumes that the speckle pattern undergoes a rectilinear translation, while here the speckle trajectory is circular, since the drift is due to the rotation of the upper plate. However, for ROIs much smaller than their distance to the center of rotation and for small enough drifts the circular motion can be safely approximated by a rectilinear drift. We find that this is indeed the case in all the situations of practical interest.

%%%%%%%%%%%%%%%%%%%%%%%%%%%%%%%%%%%%%%%%%%%%%%%%%%%%%%%%%%%%%%
\begin{figure}
\includegraphics{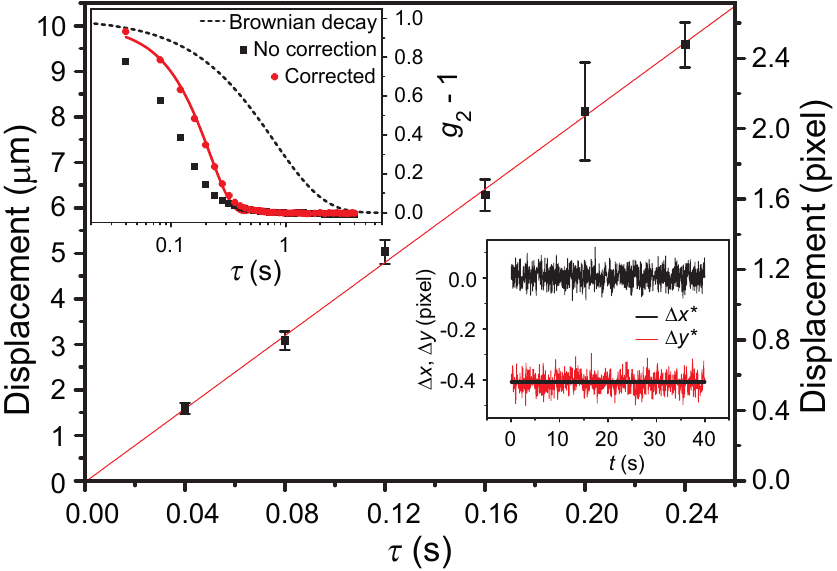}
\caption{\label{Fig5} Main plot: rigid displacement of the speckle pattern for a ROI centered on the $x$ axis, for a diluted Brownian suspension sheared at a constant rate, as a function of the lag $\tau$ between pairs of images. The error bars are the rms temporal fluctuations of the displacement. The line is a linear fit through the origin, yielding a local drift velocity of $40\pm1 \um/\mathrm{s}$. Top-left inset: raw intensity correlation function (black squares) and $g_2-1$ corrected for the drift contribution (red circles), both measured while shearing the sample. The dashed black line is the expected $g_2-1$ at rest, the red solid line is a fit to the data using Eq.~\ref{eq:g2_shear_brownian}, to account for the contribution of the affine displacements. Bottom-right inset: time dependence of the $x$ and $y$ components of the ROI displacement over a lag $\tau = 0.04~\mathrm{s}$. The black thick line shows the expected value for $\Delta y^*$.}
\end{figure}
%%%%%%%%%%%%%%%%%%%%%%%%%%%%%%%%%%%%%%%%%%%%%%%%%%%%%%%%%%%%%%%

The bottom-right inset of Fig.~\ref{Fig5} shows the $x$ and $y$ components of the speckle drift over a fixed time lag $\tau = 0.04~\mathrm{s}$, as a function of time $t$. The sample is a Brownian suspension sheared by rotating the upper plate at an angular velocity $\omega = 6.28 \times$ 10$^{-3}$ rad/s, corresponding to a tangential velocity of 75$\pm$6 $\mu$m/s (20.4 pixel/s) for the ROI used for the data analysis (the ROI position is similar to that of ROI 1 in Fig.~\ref{Fig4}). As expected for a ROI centered on the $x$ axis, $\Delta x^*$ is zero, to within random fluctuations due to the noise of the speckle tracking algorithm. The $y$ component of the displacement is in excellent agreement with the value expected from the imposed rotation speed and the location of the ROI, shown by the horizontal black line. To within the experimental noise, both $\Delta x^*$ and $\Delta y^*$ are constant, in agreement with the fact that the imposed shear rate is fixed. For stationary flow, the displacement can be more accurately measured by averaging over time. The result is shown in the main graph of Fig.~\ref{Fig5}, for six values of $\tau$. The $t$-averaged displacement grows linearly with $\tau$, as expected for uniform motion. Moreover, a linear fit through the origin of the data yields a drift velocity of $40\pm1 \um/\mathrm{s}$, in excellent agreement with $38\pm3\um/\mathrm{s}$, the drift velocity averaged over the gap as measured by the rheometer. This demonstrates the ability of our setup to measure the local, $z$-averaged flow velocity, a valuable piece of information in order to detect any deviation from uniform shear, e.g. due to wall slippage or shear banding.

The top-left inset of Fig.~\ref{Fig5} reports the intensity correlation function measured under shear. The black squares show $g_2-1$ with no drift correction. The decay time is of the order of 0.1 s. Since over that time the speckles drift by about 2 pixels, comparable to or even larger than the speckle size, we expect the drift contribution to be relevant. This is confirmed by comparing the black squares to the red circles, which show $g_2-1$ corrected for the drift, i.e. obtained by averaging $c^*_I(t,\tau)$ over time $t$. The decay of the corrected $g_2-1$ is indeed much slower. The red line is a fit to the corrected data using Eq.~\ref{eq:g2_shear_brownian}. In the fit, all parameters are fixed to their nominal values, except $\beta$ that is allowed to vary in order to account for the ROI finite size in the simple way discussed in reference to Fig.~\ref{Fig4}. Using $\beta = 3^\circ$, an excellent agreement is obtained, showing that the relative motion of the scatterers can be precisely quantified even when the shear rate is large enough to make the drift contribution significant. The inset shows also the theoretical $g_2-1$ for the same sample at rest, calculated using the nominal values $D$ = 1.5 $\times$ 10$^{-3}$ $\mu$m$^2$ s$^{-1}$ and $q$ = 33.2 $\mu$m$^{-1}$: consistently with the large value of the applied shear rate, we find that the relative motion is dominated by the affine contribution, rather than by Brownian diffusion.

\section{\label{sec:level4}Startup shear of a polymer gel}

%%%%%%%%%%%%%%%%%%%%%%%%%%%%%%%%%%%%%%%%%%%%%%%%%%%%%%%%%%%%%%%%%%%%%%%%
\begin{figure}[h!]
\includegraphics{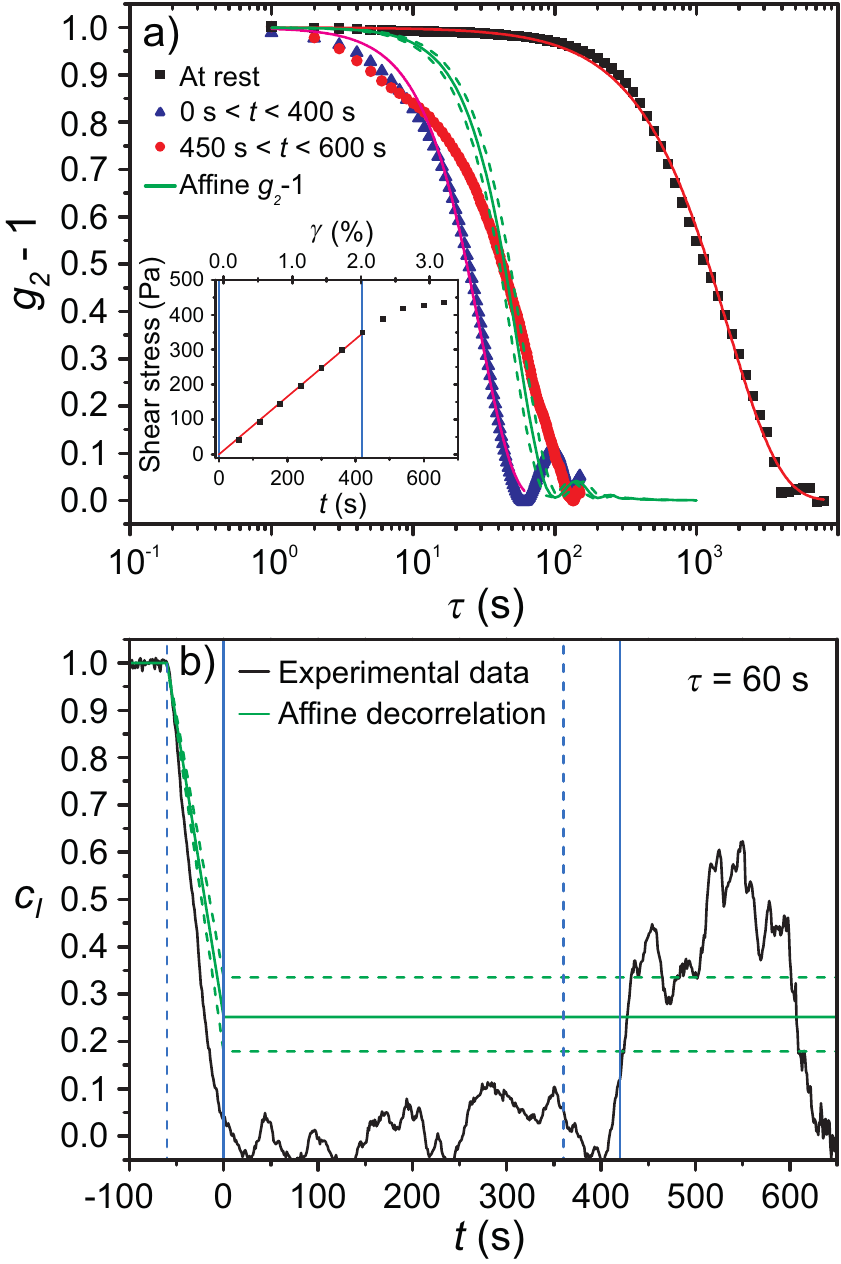}
\caption{\label{Fig6}a) Intensity correlation functions for an agarose gel under various conditions. Black squares: dynamics at rest. Blue triangles and red circles: $g_2-1$ measured during shear startup, at a constant $\dot{\gamma} = 5 \times 10^{-5}~\mathrm{s}^{-1}$, and averaged over $0~\mathrm{s} \le t \le 400~\mathrm{s}$ and $450~\mathrm{s} \le t \le 600~\mathrm{s}$, respectively. The green solid and dotted lines show the expected decay of $g_2-1$ due to affine displacements, with its confidence band. The other lines are fit to $g_2-1$ using Eq.~\ref{eq:gel}. Inset: time and strain dependence of the shear stress. The red line is a linear fit through the origin. The vertical blue line shows the end of the linear regime. b) Degree of correlation at a fixed lag $\tau = 60~\mathrm{s}$, as a function of time, with $t=0$ the shear startup time. The green lines show the expected value of $c_I$ and its confidence band if only affine displacements were present. The blue solid lines mark the start and end of the linear regime. The dashed lines indicate when the two-time degree of correlation starts to be sensitive to the change of regime indicated by the solid vertical line positioned 60 s later.}
\end{figure}
%%%%%%%%%%%%%%%%%%%%%%%%%%%%%%%%%%%%%%%%%%%%%%%%%%%%%%%%%%%%%%%%%%%%%%%%%%%%

To demonstrate the potentiality of our setup, we investigate a more complex system, measuring simultaneously the microscopic dynamics and the shear stress for an agarose gel~\cite{rochas_thermally_1994} during shear startup. The gel is prepared by mixing the agarose powder (Sigma Aldrich A9539-10G, 1\% by weight) with MilliQ water at room temperature. The solution is then heated and kept at $ T =95^{\circ}$C for 15 minutes to allow for complete agarose dissolution. The solution is poured between the rheometer plates, which are pre-heated at 95$^{\circ}$C and spaced by a gap $H= 1.025~\mathrm{mm}$. Silicon oil is put around the plates to prevent water evaporation and the temperature is cooled down to $23^{\circ}$C to form the gel \textit{in situ}. The gel is weakly scattering, such that the experiments are performed in the single scattering regime. We investigate the spontaneous dynamics of the gel (no applied shear), as well as the flow-induced dynamics upon imposing a constant shear rate of $5 \times 10^{-5}~\mathrm{s}^{-1}$.

The symbols in Fig.~\ref{Fig6}a show the intensity correlation functions obtained for a ROI with $\beta=40^\circ$ and $r = 15.2~\mathrm{mm}$, corresponding to $q_x = -3.06 \um^{-1}$ and $q_y = 0.813 \um^{-1}$. This ROI is chosen on purpose far from the $x$ axis, so as to be able to measure the contribution of the affine deformation to the decay of $g_2-1$. We fit the intensity correlation function with the general form
\begin{equation}
\begin{split}
\label{eq:gel}
g_2(\vect{q},\dot{\gamma},\tau ) - 1 & = \text{sinc}^2\left[\frac{\dot{\gamma}(x,y) \tau H} {2}(q_x \sin \beta + q_y \cos \beta)\right]\\
&  \times  \exp\left[ - (\tau / t_1)^p\right] \,,
\end{split}
\end{equation}
where the first term of the r.h.s. accounts for the contribution of the affine deformation, as in Eq.~\ref{eq:g2_shear_brownian}, while the second one is a conveniently simple form describing any other source of microscopic dynamics. The black squares are the dynamics at rest: $g_2-1$ is well fitted by Eq.~\ref{eq:gel} with $\mathrm{sinc}^2 = 1$ (because $\dot{\gamma}=0$), and a slow, compressed exponential decay ($t_1 = 1660~\mathrm{s}$ and $p=1.16 > 1$). These dynamics are consistent with the very slow compressed exponential relaxations reported for a variety of gels, with an exponent $p$ ranging from 1.5 at low $q$ down to $p \gtrsim 1$ at large $q$~\cite{cipelletti_universal_2000,duri_length_2006,guo_gel_2011,secchi_biopolymer_2013,larobina_hierarchical_2013,buzzaccaro_spatially_2015,philippe_mucus_2017}. These dynamics have been attributed to the relaxation of internal stress built up at gelation~\cite{cipelletti_universal_2000,bouchaud_anomalous_2001}.

Upon shear startup, the dynamics accelerate dramatically, as shown by the blue and red symbols in Fig.~\ref{Fig6}a. Interestingly, the dynamics significantly evolve over time, although the imposed shear rate is constant. Indeed, the intensity correlation function averaged over the first 400 s of the experiment (blue triangles) exhibits  a decay rate about twice as fast as that averaged over $450~\mathrm{s} \le t \le 600~\mathrm{s}$ (red circles). Remarkably, the change of dynamical regime around 400-450 s corresponds to the end of the linear regime where the shear stress is proportional to $t$, and thus to $\gamma$, delimited by the blue vertical lines in the inset of Fig.~\ref{Fig6}a.

In order to understand the nature of the microscopic dynamics under shear, it is important to quantify the contribution due to a purely affine deformation. We calculate this contribution by integrating the $\mathrm{sinc}^2$ term of Eq.~\ref{eq:gel} over the ROI area, using for all parameters their nominal values. This yields the green line in Fig.~\ref{Fig6}a, with the dotted lines indicating the confidence band due to the uncertainty on $q_x$ and $q_y$.
%As for the experiment reported in Fig.~\ref{Fig4}, we find that the $\mathrm{sinc}^2$ function integrated over the ROI area is very well approximated by a single $\mathrm{sinc}^2$ term with an effective value of $\beta$ slightly different from that of the ROI center ($38.15^\circ$ instead of  $40^\circ$).
In the initial regime, $t \le 400$ s and $\gamma \le 2\%$, the dynamics are faster than those expected for a purely affine deformation. This indicates that, although the mechanical gel response is linear, non-affine displacements must be present in addition to the affine deformation. We fit $g_2-1$ to Eq.~\ref{eq:gel}, fixing the contribution due to the affine deformation to that calculated by integrating over the ROI (green line in Fig.~\ref{Fig6}a) and using $t_1$ and $p$ as free parameters. The pink line shows the best fit, with $t_1 = 53.8~\mathrm{s}$ and $p = 1.71$. Except for $\tau \lesssim 20~\mathrm{s}$, where the fit slightly underestimates the decay of the correlation function, Eq.~\ref{eq:gel} reproduces very well the relaxation of $g_2-1$. The fitted decay time is 30 times shorter than for the gel at rest and about two times shorter than the contribution due to affine deformation alone. The stretching exponent $p$ is larger than one, and is larger than that at rest.

One possible explanation of the observed dynamics is that the external stress accumulated as the sample is sheared triggers rearrangement events similar to those observed %, albeit at a much lower rate,
in the sample at rest, where they originate from internal stress. Another possibility is that the non-affine dynamics are not due to irreversible plastic restructuring, but rather to deviations in the local response of the gel to that of a homogeneously elastic material, due to the heterogeneous structure of the gel. Indeed, shear-induced, non-affine microscopic dynamics observed in the small-$\gamma$ regime of other gels~\cite{basu11,colombo_stress_2014,aime18} have been attributed to spatial fluctuations of the elastic modulus~\cite{didonna_nonaffine_2005,zaccone_elastic_2013a}, rather than to plastic rearrangements.

Further insight in these dynamics may be gained by inspecting their time evolution. We find that the non-affine dynamics of our gel exhibit strong temporal fluctuations. This is seen in Fig.~\ref{Fig6}b, where we show the degree of correlation $c_I$ between speckle images taken at time $t$ and $t+\tau$ (see Eq.~\ref{eq:cI}), for a fixed time lag $\tau = 60~\mathrm{s}$ and as a function of $t$, $t=0$ being the shear startup time. The green solid line is the behavior expected for a purely affine deformation, with no additional dynamics. In this case, $c_I$ would be equal to one for $t \le -60~\mathrm{s}$, because both images are taken while the sample is at rest. The degree of correlation would progressively decrease for $-60~\mathrm{s} < t < 0$, because the first image of the pair is taken while the sample is at rest, while the second one corresponds to an increasingly deformed sample configuration. Finally, for $t \ge 0 ~\mathrm{s}$, $c_I$ would reach a steady-state value, since the sample deformation over the time lag separating the two images is constant. The experimental $c_I$ is shown by the black line: it strongly deviates from the ideal behavior of purely affine dynamics. For $t \le 400~\mathrm{s}$ $c_I$ is significantly lower than the value expected for affine deformation, indicating additional non-affine motion, as already inferred from the average dynamics. However, the degree of correlation strongly fluctuates, revealing bursts of enhanced dynamics (lower $c_I$) that last several tens of seconds. These fluctuations are not due to noise measurement, whose typical rms amplitude is much smaller, of order 0.01, as seen from the $c_I$ trace before shear startup (Fig.~\ref{Fig6}b, $t \le -60~\mathrm{s}$). They reveal the discontinuous nature of the microscopic dynamics under shear and may hint at discrete events similar to those responsible for the `avalanches' seen in simulations and experiments on driven amorphous, glassy materials~\cite{lemaitre_rate-dependent_2009,krisponeit_crossover_2014-1,antonaglia_bulk_2014},  and polymeric and colloidal gels~\cite{coussot02,kurokawa15}. Note that the experiments of Refs.~\cite{krisponeit_crossover_2014-1,antonaglia_bulk_2014} reported signatures of a non-monotonic evolution of macroscopic quantities (the stress or the deformation). Here, for $t\le 400~\mathrm{s}$ the stress evolution does not exhibit any deviation from a monotonic, linear behavior, see the inset of Fig.~\ref{Fig6}a. Most likely, this is due to the size of our sample, which is much larger than the system size in Refs.~\cite{lemaitre_rate-dependent_2009,krisponeit_crossover_2014-1,antonaglia_bulk_2014}, thereby averaging out more effectively the macroscopic quantities measured by the rheometer. This highlights the interest of space-resolved measurement such as those afforded by dynamic light scattering in the imaging geometry.

For $450~\mathrm{s} \le t \le 600~\mathrm{s}$, the degree of correlation is significantly higher, surprisingly exceeding the value expected for the affine contribution. This indicates that on the probed time scale the particle displacement is smaller than the affine component, i.e. the gel is actually deformed \textit{less} than what expected from the applied shear rate, at least in the probed ROI. Inspection of the full correlation function averaged over the same time interval (red circles in Fig.~\ref{Fig6}a) corroborates the finding that the microscopic dynamics strongly deviate from those expected for affine deformation. The decay of $g_2-1$ is less steep than that of the $\mathrm{sinc}^2$ term: at short $\tau \lesssim 20~\mathrm{s}$ it is close to that measured at the beginning of the shear startup, but at larger $\tau$ $g_2-1$ lays \textit{above} the green line calculated for affine motion, thus confirming smaller-than-expected particle displacements.
%\cite{colombo_stress_2014} sees shear banding but at $\gamma > 40\%$, much larger than the range of $\gamma$ probed here.

The only possible explanation for the behavior observed for $t \ge 450~\mathrm{s}$ is the existence of shear localisation, most likely due to wall slip, such that a sizeable fraction of the gel is indeed deformed less than for homogeneous, affine flow. To test this hypothesis, we compare the $z$-averaged ROI drift to that expected for an affine deformation. This comparison cannot be performed on the same ROI as in Fig.~\ref{Fig6}. Indeed, that ROI was chosen far from the $x$ axis, in order to make the microscopic dynamics easily measurable. As a result, $g_2-1$ decays relatively rapidly, on a time scale of at most $100~\mathrm{s}$. The expected ROI drift over $100~\mathrm{s}$ is a mere 0.05 pixel. Thus, the speckle pattern fluctuates too rapidly for the drift to be measurable. In order to make the contribution of affine deformation vanish, we need to analyze a ROI on the $x$ axis, for which the $\mathrm{sinc}^2$ term in Eq.~\ref{eq:gel} is one, the decay of $g_2-1$ much slower, and the drift measurable.

%%%%%%%%%%%%%%%%%%%%%%%%%%%%%%%%%%%%%%%%%%%%%%%%%%%%%%%%%%%%%%%%%%%%%%%%
\begin{figure}[h!]
\includegraphics{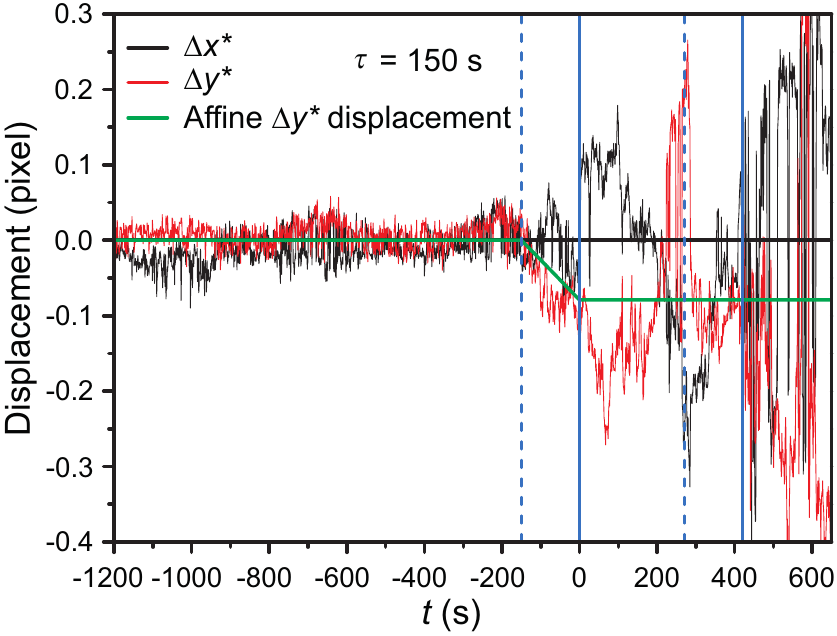}
\caption{\label{Fig7} $x$ and $y$ components of the displacement of a ROI located on the $x$ axis, for the same shear startup experiment as in Fig.~\ref{Fig6}. The blue solid vertical lines indicate the time at which shear starts and the end of the rheological linear regime, respectively. The dashed vertical lines indicate when the displacement calculated between pairs of images taken at time $t$ and $t+\tau$ starts to be sensitive to the change of rheological regime (beginning of the shear and end on linear regime). The black and green thick lines are the expected behavior of $\Delta x^*$ and $\Delta y^*$ for a purely affine deformation.}
\end{figure}
%%%%%%%%%%%%%%%%%%%%%%%%%%%%%%%%%%%%%%%%%%%%%%%%%%%%%%%%%%%%%%%%%%%%%%%%%%%%%

Figure~\ref{Fig7} shows the drift for such a ROI, positioned on the $x$ axis ($\beta = 0$, $q_y = 0 \um^{-1}$), at the same distance $r = 15.2~\mathrm{mm}$ from the tool center as the ROI of Fig.~\ref{Fig6}, so that the locally imposed shear rate is the same. The black and red lines are the $x$ and $y$ components of the drift over $\tau = 150~\mathrm{s}$, one of the shortest delays for which the ROI displacement can be reliably measured. The solid vertical blue lines mark the beginning and the end of the linear regime, as in Fig.~\ref{Fig6}. The dotted lines are positioned 150 s before the solid lines and indicate the time at which the the two-time displacement starts to be sensitive to the change of regime marked by the following vertical solid line. The green line shows the displacement $\Delta y^*$ calculated for a purely affine deformation field. For $t \le -150~\mathrm{s}$, one expects $\Delta y^* = 0$, since the displacement is calculated for pairs of images of the sample at rest. For $ t > -150~\mathrm{s}$ $\Delta y^*$ should evolve linearly towards its steady state value, since the second image of the pair is taken for an increasingly deformed sample configuration. The steady state should be reached at $t = 0~\mathrm{s}$, beyond which both images used to measure the displacement are taken while shearing the sample at a fixed rate. Throughout the experiment, $\Delta x^*$ should be null, because $\beta = q_y = 0$.

At the very beginning of shear startup, $\Delta x^*$ and $\Delta y^*$ appear to be consistent with affine deformation, as seen by comparing their trend to the expected one in the range $-150~\mathrm{s} \le t \le 0~\mathrm{s}$. However, larger-than-expected displacements for both components are seen already for $t \ge 0~\mathrm{s}$, in a regime where the macroscopic rheology is still linear. Note that in this regime the microscopic dynamics exhibit fluctuations and deviate from a purely affine deformation (see Fig~\ref{Fig6}b). Large fluctuations of both $\Delta x^*$ and $\Delta y^*$ are seen for $t \gtrsim 400~\mathrm{s}$. In this regime, not only is the behavior of the displacement inconsistent with an ideal affine deformation, it is also incompatible with simple plug flow, which could be expected, e.g., if a neat fracture was formed in the horizontal plane. Indeed, in this case the largest measurable displacement would correspond to a gel detached from the bottom, static plane and attached to the rotating top plane. For the ROI investigated here, this would correspond to $\Delta x^* = 0$ and $\Delta y^*$ twice as large as for ideal affine deformation. The strongly intermittent behavior seen in Fig.~\ref{Fig7}, with displacement values well in excess to those predicted for simple plug flow, rather point to a chaotic flow behavior, where bursts of motion relax the stress progressively built by the continuous rotation of the upper plate. These events must involve the correlated motion of large portions of the sample, in a process intriguingly reminiscent of stick-slip. Indeed, Fig.~\ref{Fig6} shows that in this regime the \textit{relative} motion of the gel network is small. Collectively, the space- and time-resolved data of Figs.~\ref{Fig6},~\ref{Fig7} show that from the very onset of shear startup the local deformation and the microscopic dynamics strongly deviate from affine deformation. The large temporal fluctuations of both the microscopic dynamics and the mesoscopic displacement suggest that these deviations are due to a series of rearrangement events associated with shear localization, rather than to the continuous (non-affine) deformation of a pristine sample. The proliferation of these events for $t \gtrsim 400~\mathrm{s}$ explains the softening of the gel measured by rheology.

As a concluding remark, it is worth mentioning that the detection of shear localization would not have been easy, based only on the rheology data. Indeed, from a rheological point of view, the change of slope seen in the inset of  Fig.~\ref{Fig6}a could also be due to extensive plastic activity that softens the gel. The absence of a stress overshoot, usually seen in shear startup of gels~\cite{colombo_stress_2014,park_structurerheology_2017,johnson_yield_2018,kurokawa15}, might suggests wall slip. However, the only way to unambiguously discriminate between the various possible origins of the shear \textit{vs} strain behavior of Fig.~\ref{Fig6}a would be to repeat the measurement using different gaps $H$, since any gap dependence of the mechanical response would point to wall slip or banding. By contrast, our setup allows an easy and quick detection of heterogeneous flow, a great advantage especially for the non-linear regime of complex systems, where reproducibility may be an issue.

\section{\label{sec:level5}Conclusion}
We have presented a home-made dynamic light scattering setup coupled to a commercial rheometer. Thanks to the use of a 2D detector in an imaging geometry, the setup allows one to measure the microscopic dynamics with both spatial and temporal resolution. Although the scattering vector is mainly oriented along the vertical, or flow gradient, direction, $\mathbf{q}$ has also a small component in the flow-vorticity plane. As a consequence, the intensity correlation function contains a contribution due to the affine deformation. We have shown that this contribution can be reliably calculated and factored out from the experimental $g_2-1$. In other apparatuses that measure DLS or XPCS under flow~\cite{busch_dynamics_2008,aimeDLSshear} in the far field limit, the decay of the correlation function is accelerated by a third contribution, stemming from the fact that sample material continuously enters and leaves the scattering volume, which is therefore completely renewed after some time. In our space-resolved apparatus, this contribution is minimized by following the average displacement of a ROI, using mixed spatio-temporal correlators. Additionally, these correlators allow the local flow velocity (averaged over the gap) to be measured, a valuable tool for checking for flow heterogeneities.

In this paper, we have presented measurements in the DLS regime. Single scattering conditions are more demanding than the strong multiple scattering regime probed by DWS, since stray light reflected by the optical interfaces or scattered by any imperfection in the optics is likely to be comparable to or even more intense than light scattered by the sample. In this respect, it is mandatory to reduce as much as possible forward scattering originating from the reflection of the incident beam on the upper plate. We have proposed and demonstrated a simple way to address this issue, by gluing a ND filter to the upper plate.

The apparatus described here can also be used for space-resolved DWS in the backscattering geometry. In this case, the ND filter is not required, since light propagating through a multiply scattering medium is strongly attenuated: the intensity of the transmitted beam reaching the upper plate is then negligible. For the same reason, the adoption of a plate-plate geometry, which is compulsory for DLS, is actually not needed for DWS. One could safely trade it for a cone-and-plate geometry, more appealing from a rheological point of view, because it insures constant stress and strain throughout the sample. Note that multiple scattering intrinsically blurs the spatial resolution of the setup, because photons emerging in a given point will have propagated through a finite portion of the sample. However, the lateral size of this region is comparable to the sample thickness, which is typically of the order of 1 mm, usually comparable to or smaller than the size of the ROIs over which $g_2-1$ has to be averaged. Thus, in practice space-resolved DWS will have a resolution comparable to that of DLS.

We hope that the relative simplicity of the setup, its flexibility and the encouraging results reported here will motivate an increasingly large number of researchers to embrace the beauty and richness of DLS measurements coupled to traditional rheology.

% If in two-column mode, this environment will change to single-column format so that long equations can be displayed.
% Use only when necessary.
%\begin{widetext}
%$$\mbox{put long equation here}$$
%\end{widetext}

% If you have acknowledgments, this puts in the proper section head.
\begin{acknowledgments}
We thank S. Aime for illuminating discussions and J.-M. Fromental for help with instrumentation. This work was supported by Agence National de la Recherche (ANR) Grant ANR-14-CE32-0005-01 and Centre National d'\'{e}tudes Spatiales (CNES)
\end{acknowledgments}

% Create the reference section using BibTeX:
%\bibliography{biblio_rheoDLS}
\providecommand{\noopsort}[1]{}\providecommand{\singleletter}[1]{#1}%

\end{document}